\documentclass[12pt,draftcls,onecolumn]{IEEEtran}

\usepackage{amsmath}
\usepackage{amssymb}
\usepackage{bbm}
\usepackage{graphicx}
\usepackage{subfigure}
\usepackage{cite}
\usepackage{algorithmic}
\usepackage{amsthm}

\newtheorem{theorem}{Theorem}
\newtheorem{cor}[theorem]{Corollary}

\newtheorem{prop}[theorem]{Proposition}

\makeatother
\begin{document}

\title{On Pseudocodewords and Decision Regions of Linear Programming Decoding of HDPC Codes}


 \author{\IEEEauthorblockN{Asi~Lifshitz and Yair~Be'ery~\IEEEmembership{Senior Member,~IEEE}}
 \IEEEauthorblockA{\\ Tel Aviv University, School of Electrical Engineering \\
Ramat Aviv 69978, ISRAEL\\
 Email:  asilifsh@post.tau.ac.il, ybeery@eng.tau.ac.il}}
 
\maketitle

\begin{abstract}
  In this paper we explore the decision regions of Linear Programming (LP) decoding. We compare the decision regions of an LP decoder, a Belief Propagation (BP) decoder and the optimal Maximum Likelihood (ML) decoder. We study the effect of minimal-weight pseudocodewords on LP decoding. We present global optimization as a method for finding the minimal pseudoweight of a given code as well as the number of minimal-weight generators. We present a complete pseudoweight distribution for the $[24, 12, 8]$ extended Golay code, and provide justifications of why the pseudoweight distribution alone cannot be used for obtaining a tight upper bound on the error probability.

\end{abstract}

\section{Introduction}
\IEEEPARstart
The behavior of the BP \cite{Kschischang-Factor-graphs} decoder for the case of finite-length codes does not have simple characteristics, and can be very hard to predict. Linear programming is a well-studied discipline that provides efficient analysis tools. The relationship between linear programming decoding and belief propagation decoding was observed and characterized \cite{Vontobel-On-the-relationship}, and the decision regions of these decoders are suggested to be tightly related.

The LP decoder receives the channel likelihood ratios which define an objective function, for which it finds an optimal solution that satisfies a set of constraints. These constraints are inequalities arisen from a given parity check matrix and form a polytope, also known as the \textit{fundamental polytope} \cite{Vontobel-Graph-cover-decoding}. The fundamental polytope is a relaxation of the codewords polytope. It has a clear geometrical representation which is well-suited for finite-length analysis. The vertices of the fundamental polytope are every codeword, but also some non-codewords pseudocodewords \cite{Feldman-Using-linear-programming}. The \textit{fundamental cone} \cite{Vontobel-Graph-cover-decoding} is the conic hull of the fundamental polytope. It has a vertex in the origin, and its edges are also referred to as \textit {minimal pseudo-codewords} \cite{Vontobel-Graph-cover-decoding} or \textit {generators} \cite{Siegel-Relaxation-bounds}. The fundamental cone has a more compact representation than the fundamental polytope, and it is sufficient to consider the fundamental cone for evaluating the performance of the LP decoder \cite{Siegel-Relaxation-bounds}.

The output of the LP decoder is always a vertex of the polytope which maximizes the channel likelihood ratios function. One of the most appealing properties of the LP decoder is the \textit{ML~certificate} property - whenever it returns an integral solution, the solution is guaranteed to be the ML codeword; otherwise an error is invoked. There are rare cases \cite{Kashyap-decomposition-theory} for which the vertices of the polytope are codewords only, and in these cases the output of the LP decoder is identical to the output of the ML decoder. In these rare cases a polynomial-time ML decoding is attainable. However, for most cases, and when applied to good error-correcting codes, the LP decoder will suffer from decoding failures due to the presence of pseudocodewords.

The minimal \textit{pseudoweight} \cite{Vontobel-Graph-cover-decoding} of a pseudocodeword in LP decoding is the appropriate analog of the minimal Hamming weight in ML decoding. Furthermore, the minimum Hamming weight is known to be lower bounded by the minimal pseudoweight \cite{Vontobel-Lower-bounds}. There are cases where the minimal pseudoweight equals the minimal Hamming weight, and in these cases, the existence of pseudocodewords may have a minor or even a negligible effect on the decoder's optimality.

High Density Parity Check (HDPC) codes are characterized by a dense parity check matrix. Linear classical codes have a dense parity check matrix by design, which makes them less suitable for LP decoding. The denser the parity check matrix is the more vertices the fundamental polytope will have. Keeping in mind that the number of codewords of a given code is constant, one can realize that increasing the number of vertices is equivalent to increasing the number of pseudocodewords which are not codewords.

The BP algorithm is often in use for decoding low-density parity-check (LDPC) codes, for which it has both low complexity and good performance. Its low complexity is achieved due to the fact that the algorithm operates locally on a so-called \textit {Tanner graph} \cite{Wiberg-Codes-and-Decoding} representation of the parity-check matrix. However, operating locally leads to a fundamental weakness of the algorithm~-~it may fail to converge due to non-codewords pseudocodewords. These pseudocodewords are valid assignments of the \textit {computation tree} \cite{Wiberg-Codes-and-Decoding} of the given code and decoder. Kelley and Sridhara \cite{Kelley-Pseudocodewords-of-Tanner-graphs} have proved that the pseudocodewords of the computation tree are a superset of the pseudocodewords which lie in the aforementioned fundamental polytope.

The decision regions of a decoding algorithm provide a visualization of the decoder's decisions upon receiving channel signals. They provide a better intuition of the decoder operation, and can be used for comparing different decoding algorithms. The existence of pseudocodewords in iterative decoding and their effect on the decision regions were studied in \cite{Vontobel-Graph-cover-decoding}, \cite{Frey-Signal-space-characterization}. In the present work we examine the effect of pseudocodewords on the decision regions and on the performance of the LP decoder.

Presenting a complete picture of the decision regions is usually impossible even for short-length codes, due to the number of dimensions involved in each decoded signal. Nevertheless, performing cuts in the signal space can provide a clear picture of specific decision regions, which illustrate the effect of pseudocodewords on the performance of the BP and LP decoders.

In order to illustrate the different decision regions of the BP, LP and ML decoders, the [8,~4,~4]  extended Hamming code was chosen. It is known that both BP and LP decoders are affected by the selection of the parity check matrix, therefore three different representations for the aforementioned code were chosen.

The contribution of this paper is in providing a better understanding of the LP decoder operation. Global optimization is proposed as a method for finding the minimal weight generators of a given parity check matrix. The paper also presents the difficulties in obtaining a tight LP union bound, based on the generators' weight distribution, and explores the effect of minimal-weight pseudocodewords on the decision regions of the LP decoder.

The rest of the paper is organized as follows: We provide some background on decision regions and a method to produce the decision regions in Section \ref{sec:mapping of decision regions}. In Section \ref{sec:the affect of minimal weight pseudocodewords on lp decoding} we present the different effect of minimal-weight pseudocodewords on LP decoding. A global optimization approach for finding the minimal weight generators is described in Section \ref{sec:search for low weight pseudocodeword}. In Section \ref{sec:LP Union Bound} we presents an LP union bound based on the generators' weight distribution. Section \ref{sec:Conclusion} concludes the paper.

\hfill
\section{mapping of decision regions}
\label{sec:mapping of decision regions}
The major difficulty of presenting the decision regions of a code longer than three is how to project or reduce an
$n$-dimensional space to a two or three dimensional subspace. In this paper the $n$-dimensional space is sliced to a two-dimensional Euclidian subspace. A two dimensional subspace or a \textit {cut} is a plane that is spanned by two orthogonal vectors.

Consider transmitting an $n$-dimensional signal over an AWGN channel, such that the observed data is $\mathbf{r} = \mathbf{s} + \mathbf{n}$ where $\mathbf{s}$ is the transmitted signal, and $\mathbf{n}$ is a normally distributed noise with zero mean and variance $\sigma^2$.
The decision regions $\left\{Z_1,\ldots,Z_M \right\}$ are the subsets of the signal space $\mathbb{R}^n$ defined by
\begin{equation}
  \label{eq1}
  Z_i=\left\{\mathbf{r}:p(\mathbf{s}_i | \mathbf{r})>p(\mathbf{s}_j |\mathbf{r})\;\; \forall i\not=j, \; 1 \leq i, j \leq M \right\}
\end{equation}
where $M$ is the number of codewords.

The decision boundaries are all the points for which exists $\mathbf{r}\in \mathbb{R}^n$ such that $p(\mathbf{s}_i|\mathbf{r}) = p(\mathbf{s}_j|\mathbf{r})$. The decision boundaries divide the signal space into $M$ disjoint decision regions, each of which consists of all the point in $\mathbb{R}^n$ closest in Euclidian distance to the received signal $\mathbf{r}$.
An ML decoder finds which decision region $Z_i$ contains $\mathbf{r}$, and outputs the corresponding codeword $\hat{\mathbf{c}_i}$. The existence of pseudocodewords in BP and LP decoders divides the signal space into more decision regions than those created solely by codewords. Clearly, these pseudocodewords reduce the decision regions of the codewords, hence deteriorate the decoder optimality.

The first step towards mapping the decision regions is to decide of the two spanning vectors $\mathbf{n}_y$ and $\mathbf{n}_x$ $(\mathbf{n}_y, \mathbf{n}_x \in \mathbb{R} ^n)$. These two vectors must be orthogonal in order to have a clear 2-dimensional picture, rather than a folded one. The sum of these two vectors is the noise vector that is added to the transmitted signal. The noisy samples are then decoded, and the output of the decoder is recorded along with the received signal. All the received signals which share the same output designate a decision region.
In the following the LP decoder uses a $\mathcal{C}-symmetric$ polytope \cite{Feldman-Using-linear-programming} under a binary-input memoryless symmetric channel, thus one may assume without loss of generality that the all-zero codeword was transmitted. For BP and ML decoders, the conditional decoding error probability is independent of the codeword that was sent. Therefore, our analysis will assume that the all-zero codeword was transmitted over an AWGN channel using a BPSK modulation. The BP decoder being used is a sum-product decoder, configured to perform 50 decoding iterations.

In LP decoding, the vertices of the fundamental polytope are a superset of the codewords. While the set of codewords are dominated by the code itself, the set of pseudocodewords is dominated by the relaxation being used.
Let $\mathcal{C}$ be a binary code $\mathcal{C} \in \{0,1\}^n$ and let $\mathcal{V(\mathcal{P})}$ be the set of vertices of the polytope $\mathcal{P}$. The polytope contains every codewords, but also some fractional pseudocodewords, thus:
\begin{equation}
  \label{eq4}
  \mathcal{C} \subseteq \mathcal{V(\mathcal{P})} \subseteq \mathcal{P} \subseteq [0,1]^n.
\end{equation}

The mapping of a vertex onto a point in a Euclidian plane is performed using the \textit{effective squared Euclidian distance}  \cite{Forney-On-the-effective-weights} between a codeword $\mathbf{c}$ and a pseudocodeword $\mathbf{p}$ in a balanced computation tree:

\begin{equation}
  \label{eq5}
  d_{eff}^2({\mathbf{c}, \mathbf{p}})=\frac{(\Vert \mathbf{d}^2 \Vert + \sigma_p^2)^2}{\Vert \mathbf{d}^2 \Vert}
\end{equation}
where $\mathbf{d} = \mathbf{c} - E[\mathbf{p}]$ and $\sigma_p^2 = E[\Vert \mathbf{p} \Vert ^2 ] - \Vert E[\mathbf{p}] \Vert ^2$.
If the all-zero codeword is transmitted using a BPSK modulation, then (\ref{eq5}) is simplified to

\begin{equation}
  \label{eq6}
  d_{eff}^2({\mathbf{0}, \mathbf{p}})=4w_{eff}(\mathbf{p})
\end{equation}
where $w_{eff}$ is the effective Hamming weight in an AWGN channel, given by

\begin{equation}
  \label{eq7}
  w_p^{AWGNC}(\mathbf{p})=w_{eff}(\mathbf{p}) \equiv \frac {\Vert \mathbf{p} \Vert _1 ^2}{\Vert \mathbf{p} \Vert _2 ^2}=\frac {(\sum_{i=1}^{n}p_i)^2}{\sum_{i=1}^{n}p_i^2}.
\end{equation}
Eq. (\ref{eq7}) is sometimes referred to as the \emph{pseudoweight} \cite{Vontobel-Graph-cover-decoding} of $\mathbf{p}$ in an AWGN channel. The performance of iterative decoders is influenced mostly by the minimal weight pseudocodewords \cite{Vontobel-Lower-bounds}, \cite{Chertkov-Efficient-Pseudo-Codeword-Search-Algorithm}, \cite{Vontobel-On-the-minimal-pseudo-codewords}, \cite{Chertkov-Reducing}, while the ML decoder is influenced mostly by the code minimal Hamming weight. This is why the performance of an ML decoder is not affected by the selection of the parity check matrix representation, which is not true for the case of BP and LP decoders. If all codewords are chosen with equal probability, then the effective Euclidian distance between the all-zero codeword and the decoded word is $d_{eff}$. The decision boundary between the all-zero codeword and the decoded word is exactly $\frac {d_{eff}}{2}$ from the origin.

The mapping of the decision regions within a two-dimensional cut is performed as follows:
\begin{algorithmic}[1]
  \STATE Set the y-Axis to represent $\mathbf{n}_y$, and the x-Axis to represent $\mathbf{n}_x$
  \STATE Find the normalization factor for $\mathbf{n}_y$ and $\mathbf{n}_x$:

  \begin{equation}
    \label{eq2}
    norm\_n_y=\sqrt{\sum_{i=1}^{n}(\mathbf{n}_{y_i})^2},\;
    norm\_n_x=\sqrt{\sum_{i=1}^{n}(\mathbf{n}_{x_i})^2}
  \end{equation}

  \FOR {y in range $min\_added\_noise$ to $max\_added\_noise$}
  \FOR {x in range $min\_added\_noise$ to $max\_added\_noise$}
  \STATE Set the received signal
  \begin{equation}
    \label{eq3}
    \mathbf{r}=Modulate(\mathbf{0}) + \frac{\mathbf{n}_y \cdot y}{norm\_n_y}
    + \frac{\mathbf{n}_x \cdot x}{norm\_n_x}
  \end{equation}

  \STATE Decode the signal

  \ENDFOR
  \ENDFOR
  \STATE Map the entire space spanned by y and x to the decoded words.
\end{algorithmic}

Normalization of $\mathbf{n}_y$ and $\mathbf{n}_x$ is required in order to maintain a Euclidian space. A unit-step in the direction of $\mathbf{n}_y$ is a step for which the noise that is added in the direction of $\mathbf{n}_y$ will shift the transmitted signal by $\sqrt{\sum_{i=1}^{n}(\mathbf{n}_{y_i})^2}$ from the origin.
The $min\_added\_noise$ and the $max\_added\_noise$ are the two endpoints of both $\mathbf{n}_y$ and $\mathbf{n}_x$, and are taken such that $min\_added\_noise~\ge~- \frac{d_{min}}{2}$ and $max\_added\_noise~\le~d_{min}$, where $d_{min}$ is the code minimal Hamming distance.

In the following the decision regions of the [8, 4, 4] extended Hamming code are studied. The [8, 4, 4] extended Hamming code is well-suited for studying the decision regions of the BP and LP decoders. It is a self-dual code which has a simple parity check matrix representation with minimal pseudoweight equals $d_{min}$, but also a representation with pseudoweight equals 3.  In order to gain a better understanding of the tight relation between the selected parity check matrix, the decision regions and the decoder's performance, three different parity check matrices are investigated (\ref{H1}), (\ref{H2}), (\ref{H3}). These matrices were originally introduced by Halford and Chugg in \cite{Halford-Random-Iterative}, for which they also presented the pseudoweight spectra.

\begin{equation}
  \label{H1}
  H_1=\left(
    \begin{array}{cccccccc}
      1 & 1 & 1 & 1 & 1 & 1 & 1 & 1\\
      0 & 1 & 0 & 1 & 0 & 1 & 0 & 1\\
      0 & 0 & 1 & 1 & 0 & 0 & 1 & 1\\
      0 & 0 & 0 & 0 & 1 & 1 & 1 & 1\\
    \end{array}
  \right)
\end{equation}

\begin{equation}
  \label{H2}
  H_2=\left(
    \begin{array}{cccccccc}
      1 & 1 & 1 & 1 & 0 & 0 & 0 & 0\\
      0 & 0 & 1 & 1 & 1 & 1 & 0 & 0\\
      0 & 0 & 0 & 0 & 1 & 1 & 1 & 1\\
      0 & 1 & 1 & 0 & 0 & 1 & 1 & 0\\
    \end{array}
  \right)
\end{equation}

\begin{equation}
  \label{H3}
  H_3=\left(
    \begin{array}{cccccccc}
      0 & 0 & 0 & 0 & 1 & 1 & 1 & 1 \\
      0 & 0 & 1 & 1 & 0 & 0 & 1 & 1 \\
      0 & 0 & 1 & 1 & 1 & 1 & 0 & 0 \\
      0 & 1 & 0 & 1 & 0 & 1 & 0 & 1 \\
      0 & 1 & 0 & 1 & 1 & 0 & 1 & 0 \\
      0 & 1 & 1 & 0 & 0 & 1 & 1 & 0 \\
      0 & 1 & 1 & 0 & 1 & 0 & 0 & 1 \\
      1 & 0 & 0 & 1 & 0 & 1 & 1 & 0 \\
      1 & 0 & 0 & 1 & 1 & 0 & 0 & 1 \\
      1 & 0 & 1 & 0 & 0 & 1 & 0 & 1 \\
      1 & 0 & 1 & 0 & 1 & 0 & 1 & 0 \\
      1 & 1 & 0 & 0 & 0 & 0 & 1 & 1 \\
      1 & 1 & 0 & 0 & 1 & 1 & 0 & 0 \\
      1 & 1 & 1 & 1 & 0 & 0 & 0 & 0 \\
    \end{array}
  \right)
\end{equation}

\begin{figure}
  \centering
  \subfigure[Performance]{
    \includegraphics[scale=0.6]{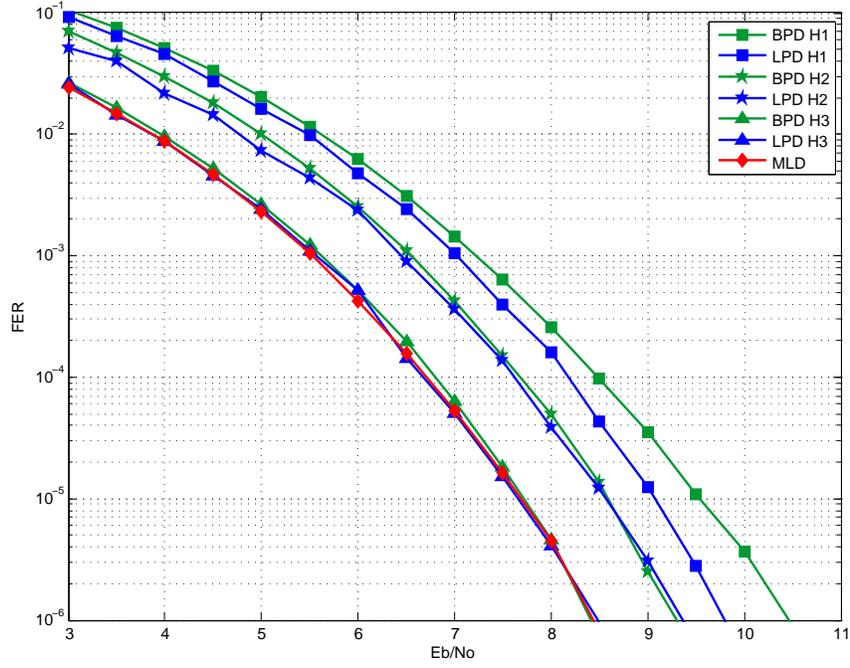}
    \label{fig:DesionRegions_FER_H1H2H3}
  }

  \subfigure[Pseudoweight spectra]{
    \includegraphics[scale=0.64]{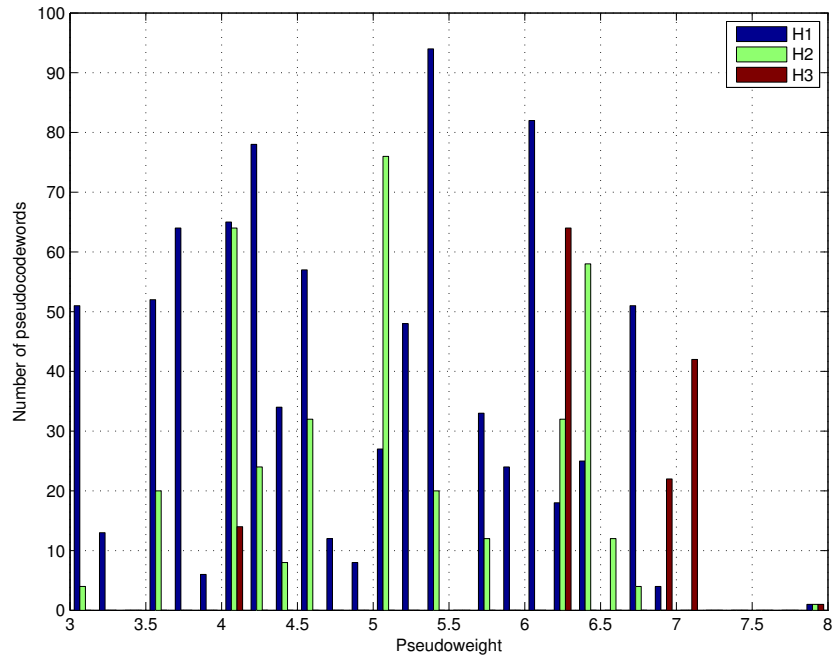}
    \label{fig:DesionRegions_H1H2H3_pcws_histogram}
  }

  \caption{Different representations of the [8, 4, 4] extended Hamming code.}
  \label{fig:DesionRegions_H1H2H3}
\end{figure}


\begin{figure} [b!]
  \centering

  \subfigure[LP decoder]{
    \includegraphics[scale=0.38]{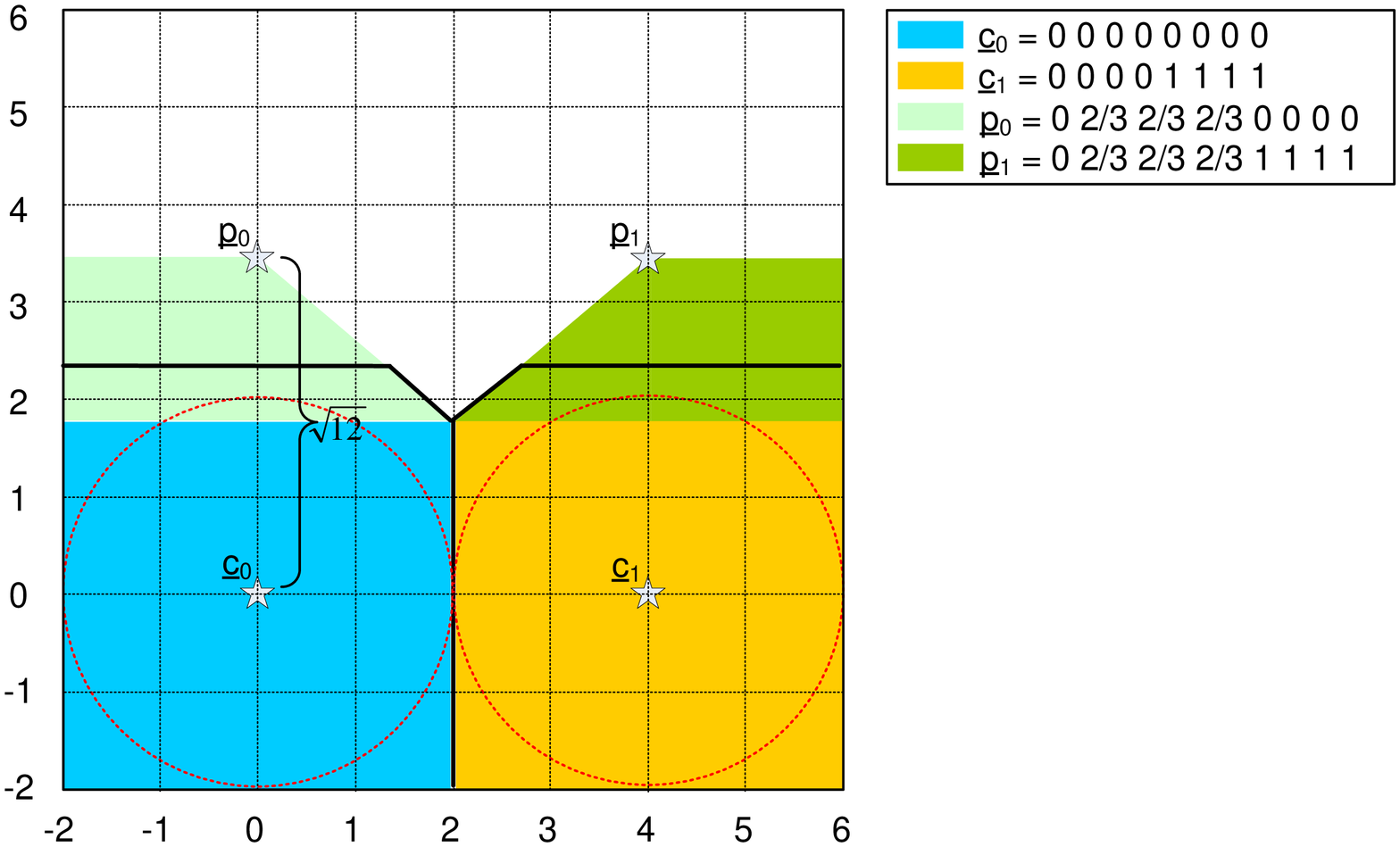}
    \label{fig:DesionRegions_H1_LP}}

  \subfigure[BP decoder]{
    \includegraphics[scale=0.38]{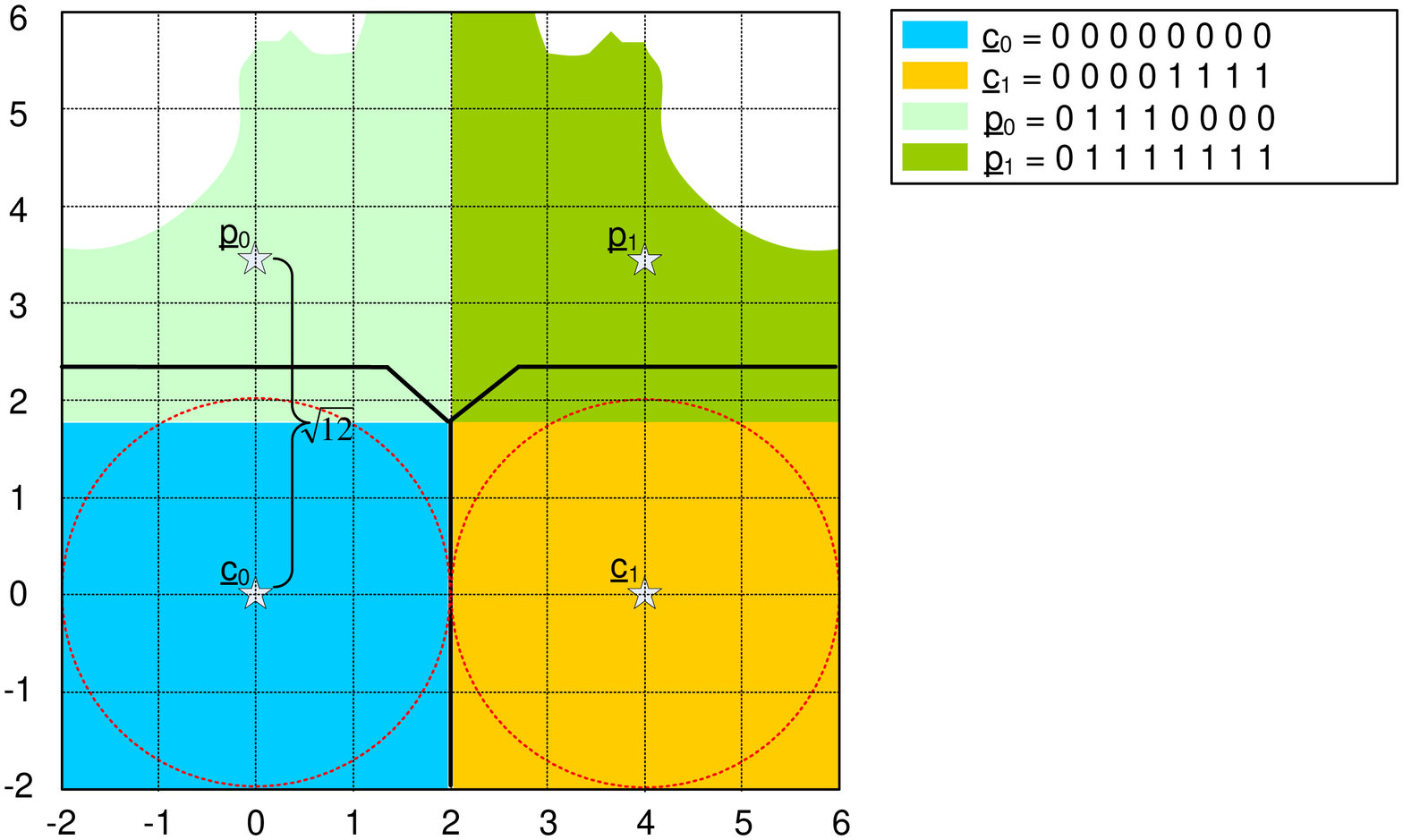}
    \label{fig:DesionRegions_H1_BP}}

  \caption{Decision regions of $H_1$ with $\mathbf{n}_y=~(0, 2/3, 2/3, 2/3, 0, 0, 0, 0)$  and  $\mathbf{n}_x=~(0, 0, 0, 0, 1, 1, 1, 1)$.}
  \label{fig:DesionRegions_H1}
\end{figure}

\begin{figure}
  \centering

  \subfigure[LP decoder]{
    \includegraphics[scale=0.38]{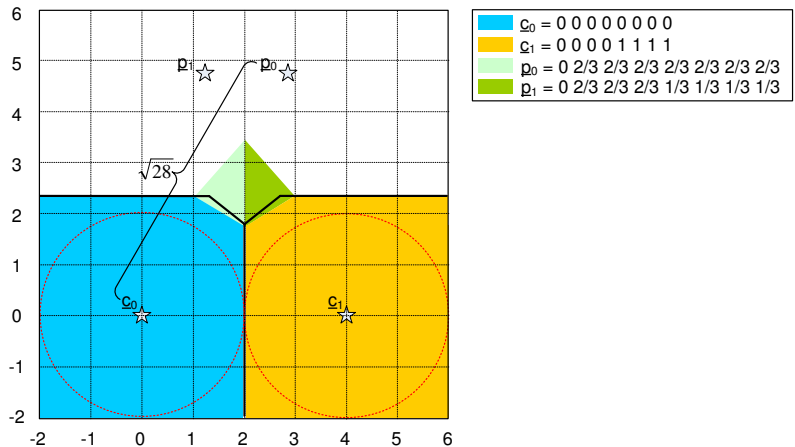}
    \label{fig:DesionRegions_H3_LP}}

  \subfigure[BP decoder]{
    \includegraphics[scale=0.38]{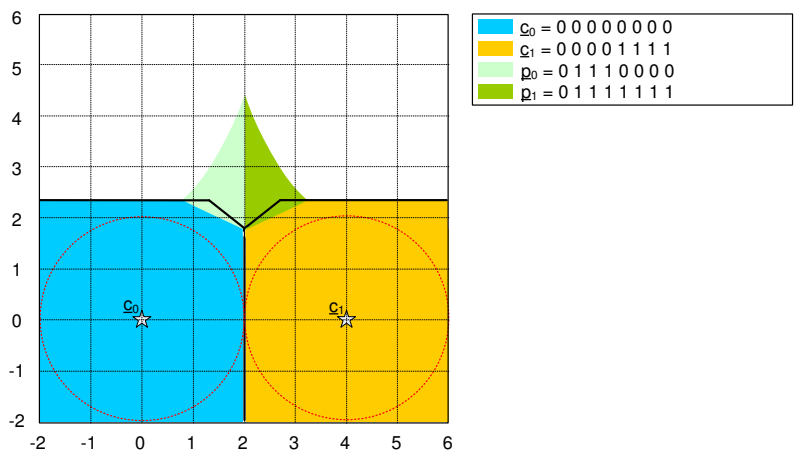}
    \label{fig:DesionRegions_H3_BP}}

  \caption{Decision regions of $H_3$ with $\mathbf{n}_y=~(0, 2/3, 2/3, 2/3, 0, 0, 0, 0)$  and  $\mathbf{n}_x=~(0, 0, 0, 0, 1, 1, 1, 1)$.}
  \label{fig:DesionRegions_H3}
\end{figure}

Fig.~\ref{fig:DesionRegions_H1H2H3} presents the performance and the weight distribution of the three representations of the [8, 4, 4] extended Hamming code. Fig.~\ref{fig:DesionRegions_FER_H1H2H3} compares the frame error rate of the LP and BP decoders using the parity check matrices of (8), (9) and (10).  The performance difference between the LP and BP decoders is consistent for the 3 representations, in which the LP decoder slightly outperforms the performance of the BP decoder.  Both decoders achieve the best performance when using $H_3$ and worst performance when using $H_1$. Furthermore, the LP decoder has almost the same performance as the ML decoder when using $H_3$.

Fig.~\ref{fig:DesionRegions_H1H2H3_pcws_histogram} illustrates the pseudoweight distribution of $H_1$, $H_2$ and $H_3$. These results are similar to the ones presented in \cite{Halford-Random-Iterative}.  A first observation is that $w_{p,min}^{AWGNC}(H_3)=d_{min}=4$, which can provide an explanation of why the suboptimal LP decoder is almost optimal. A second observation is that $H_1$ has many more low-weight pseudocodewords compares to $H_2$, which is consistent with the performance difference between the two representations.

Fig.~\ref{fig:DesionRegions_H1} and Fig.~\ref{fig:DesionRegions_H3} illustrate some of the decision regions that were found by mapping the decision regions. The solid black lines represent the optimal decision regions of the ML decoder. The bottom left decision region represents the transmitted all-zero codeword. The decision region on its right (if any) represents another codeword which is a linear combination of $\mathbf{n}_y$ and $\mathbf{n}_x$ that lies in the same plane. The output of an ML decoder can only be a codeword; hence the region above the solid lines is a region of codewords which are not a linear combination of $\mathbf{n}_y$ and $\mathbf{n}_x$. The soft output of the BP decoder enters a hard decision decoder to maintain a binary vector.

Fig.~\ref{fig:DesionRegions_H1} illustrates the decision regions in the plane spanned by  $\mathbf{n}_y=~(0, 2/3, 2/3, 2/3, 0, 0, 0, 0)$  and  $\mathbf{n}_x=~(0, 0, 0, 0, 1, 1, 1, 1)$. Clearly, the decision boundary between $\mathbf{c}_0$ and $\mathbf{c}_1$ obeys the $\frac{d_{min}}{2}$  rule. In the direction of $\mathbf{n}_y$, the decision boundary of the ML decoder is beyond $\frac{d_{min}}{2}$, since there is no competing codeword in this direction. The word $\mathbf{p}_0$ in Fig.~\ref{fig:DesionRegions_H1_LP} is a pseudocodeword, since it is fractional and has a weight equals 3 which is smaller than $d_{min}$. Fig.~\ref{fig:DesionRegions_H1_LP} also illustrates how the minimal-weight pseudocodeword $\mathbf{p}_0$ deteriorates the decoder's optimality by reducing the decision region of the transmitted codeword.

Fig.~\ref{fig:DesionRegions_H1_LP} and Fig.~\ref{fig:DesionRegions_H1_BP} show that BP and LP decoders share the same decision boundaries between $\mathbf{c}_0$ and $\mathbf{c}_1$, and same boundaries between $\mathbf{c}_0$ and $\mathbf{p}_0$. The location of $\mathbf{p}_0$ in Fig.~\ref{fig:DesionRegions_H1_LP} is $d_{eff}(\mathbf{0}, \mathbf{p}_0)=\sqrt{4w_p(\mathbf{p}_0)}=\sqrt {4 \frac{(2/3+2/3+2/3)^2}{(2/3)^2+(2/3)^2+(2/3)^2}}=\sqrt{12}$  from $\mathbf{c}_0$. The same calculation holds for Fig.~\ref{fig:DesionRegions_H1_BP}:  $d_{eff}(\mathbf{0}, \mathbf{p}_0)=\sqrt{4w_p(\mathbf{p}_0)}=\sqrt {4 \frac{(1+1+1)^2}{(1)^2+(1)^2+(1)^2}}=\sqrt{12}$. The decision boundaries are exactly at $\frac{d_{eff}}{2}=\frac{\sqrt{12}}{2}=\sqrt{3}$  from $\mathbf{c}_0$. Since the decision boundary is smaller than $\frac{d_{min}}{2}$, the LP and BP decoders are not \emph{bounded distance} \cite{Amrani-ounded-distance-decoding}, \cite{Fishler-Geometrical} decoders. The difference between BP and LP decoders is in the decision regions of pseudocodewords, and is caused due to the different algorithms that are used. While the decision regions of the LP decoder are convex polytopes \cite{Chertkov-Polytope-of-Correct}, their counterparts in BP decoding are non-convex and more chaotic. Clearly, the decision regions of $\mathbf{p}_0$ and $\mathbf{p}_1$ in BP decoding are larger than their counterparts in LP decoding. Nevertheless, from this figure it is clear that this difference has no major impact on the performance.

Fig.~\ref{fig:DesionRegions_H3} represents the decision regions when using $H_3$ and the same $\mathbf{n}_y$ and $\mathbf{n}_x$ as in Fig.~\ref{fig:DesionRegions_H1}. From Fig.~\ref{fig:DesionRegions_H3} one can observe that $H_3$ does not contain the $(0, 2/3, 2/3, 2/3, 0, 0, 0, 0)$ pseudocodeword, but rather the higher-weight pseudocodewords $(0, 2/3, 2/3, 2/3, 2/3, 2/3, 2/3, 2/3)$  and $(0, 2/3, 2/3, 2/3, 1/3, 1/3, 1/3, 1/3)$ of weight 7 and 6.25, respectively. In this case the pseudocodewords barely reduce the decision region of the transmitted codeword, which explains why $H_3$ defines a polytope which makes the LP decoder almost optimal. One can observe that $\mathbf{p}_0$ and $\mathbf{p}_1$ in Fig.~\ref{fig:DesionRegions_H3_BP} reduce the optimal decision region of $\mathbf{c}_0$ slightly more than their counterparts in Fig.~\ref{fig:DesionRegions_H3_LP}, but still maintain a bounded distance decoding. This observation is correlated with the actual performance of the two decoders. The position of BP pseudocodewords in signal space is sometimes misleading, due to the information loss caused by the hard decision at the output of the BP decoder. This is why we omitted the position of $\mathbf{p}_0$ and $\mathbf{p}_1$ from Fig.~\ref{fig:DesionRegions_H3_BP}. The fact that the decision regions of $\mathbf{p}_0$ and $\mathbf{p}_1$ in BP decoding are much larger than those of the LP decoder does not necessarily reflected in the performance, since the majority of the area is located outside the optimal decision region of $\mathbf{c}_0$, i.e. in the error region.

\section{The effect of minimal weight pseudocodewords on lp decoding}
\label{sec:the affect of minimal weight pseudocodewords on lp decoding}

\begin{figure}
  \centering

  \subfigure[$H_1$]{
    \includegraphics[scale=0.38]{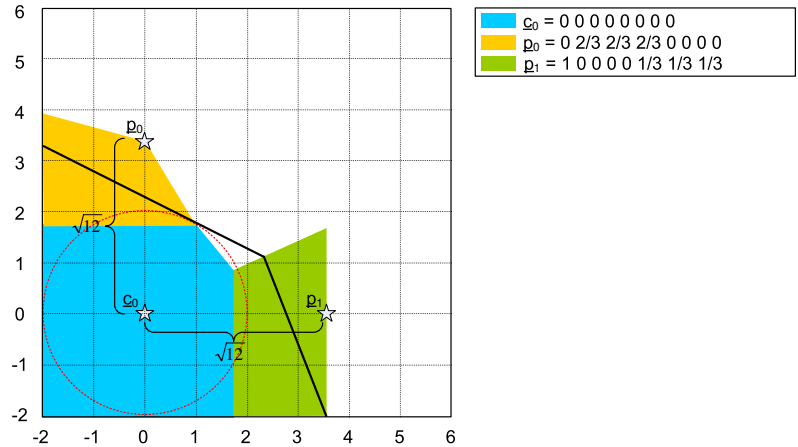}
    \label{fig:DesionRegions_H1H2H3_LP_H1}
  }

  \subfigure[$H_2$]{
    \includegraphics[scale=0.38]{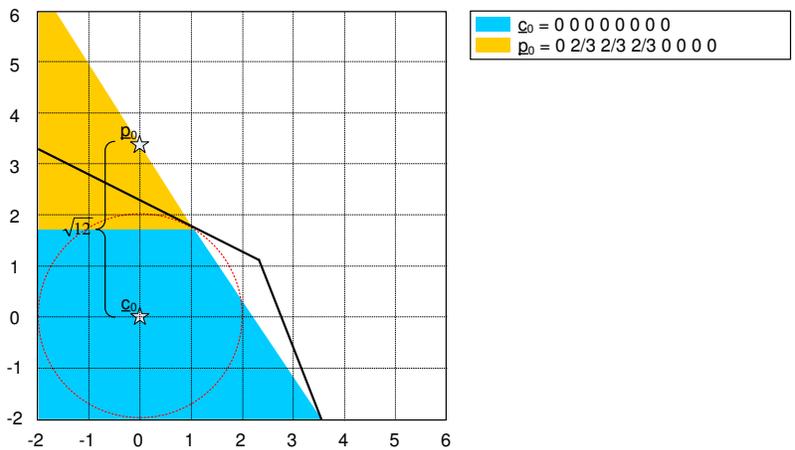}
    \label{fig:DesionRegions_H1H2H3_LP_H2}
  }
  \subfigure[$H_3$]{
    \includegraphics[scale=0.38]{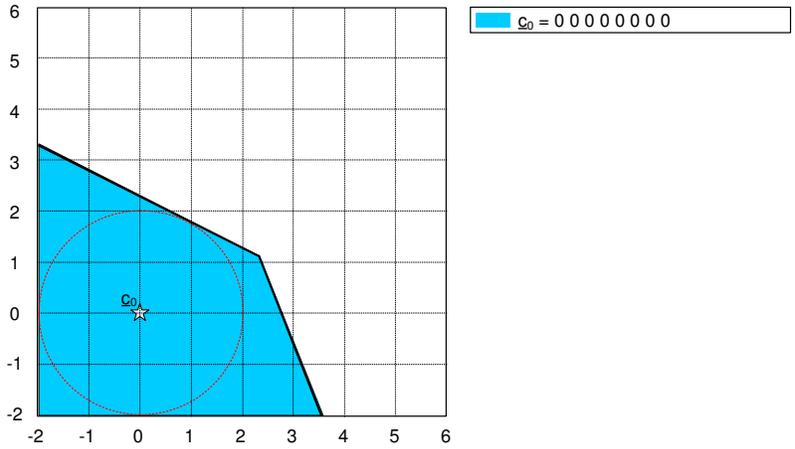}
    \label{fig:DesionRegions_H1H2H3_LP_H3}
  }

  \caption{Decision regions of an LP decoder with $\mathbf{n}_y=~(0, 2/3, 2/3, 2/3, 0, 0, 0, 0)$  and  $\mathbf{n}_x=~(1, 0, 0, 0, 0, 1/3, 1/3, 1/3)$.}
  \label{fig:DesionRegions_H1H2H3_LP}
\end{figure}

Simulations show that although the performance of the LP decoder is dominated by low-weight pseudocodewords, not all low-weight pseudocodewords have the same contribution to the error probability. In this section we will justify why some minimal weight pseudocodewords may have a higher contribution to the error probability compared to others. We will base our justification on both simulation results and decision regions.

The polytope of $H_1$ has 26 vertices of pseudoweight equals 3. Our simulation results for different SNRs show, for example, that the pseudocodeword $\mathbf{p}_0=~(0, 2/3, 2/3, 2/3, 0, 0, 0, 0)$ causes approximately 4 times more decoding errors than $\mathbf{p}_1=~(1, 0, 0, 0, 0, 1/3, 1/3, 1/3)$ . There are several properties that affect the error probability of a given pseudocodeword. When listing all pseudocodewords, one can see that there are many pseudocodewords with support equals 4 that share 3 out of 4 non-zero components with $\mathbf{p}_1$. It means that in the objective function the selection between such candidates depends on two independent random variables. However, there are no pseudocodewords with support equals 3 that share 2 components out of 3 with $\mathbf{p}_0$. There are pseudocodewords with support higher than 3 that contain non-zero components in the same positions as $\mathbf{p}_0$ but with lower values. Such components have weaker effect on the cost value, and lead to fewer decoding errors.

We will now present a cut that contains the aforementioned $\mathbf{p}_0$ and $\mathbf{p}_1$, and show that in the specific cut the decision region of $\mathbf{p}_0$ is larger than that of $\mathbf{p}_1$, which provide another perspective of why $\mathbf{p}_0$ causes more decoding errors. Fig.~\ref{fig:DesionRegions_H1H2H3_LP} presents a cut created by the noise vectors $\mathbf{n}_y=~(0, 2/3, 2/3, 2/3, 0, 0, 0, 0)$ and $\mathbf{n}_x=~(1, 0, 0, 0, 0, 1/3, 1/3, 1/3)$. Notice that the decision regions, at least in the presented cut, have a different behavior as illustrated in Fig.~\ref{fig:DesionRegions_H1H2H3_LP_H1}. While the decision region of $\mathbf{p}_1$ increases only in the y-axis, the decision region of $\mathbf{p}_0$ grows in both axes.

The existence of $\mathbf{p}_0$ and $\mathbf{p}_1$ in the polytope of $H_1$ reduces the decision region of $\mathbf{c}_0$ as shown in Fig.~\ref{fig:DesionRegions_H1H2H3_LP_H1}. The polytope of $H_2$ does not contain $\mathbf{p}_1$, thus the decision region of $\mathbf{c}_0$ is bigger than that of $H_1$ as presented in Fig.~\ref{fig:DesionRegions_H1H2H3_LP_H2}. The fundamental polytope of $H_3$ does not contain $\mathbf{p}_0$, nor $\mathbf{p}_1$, thus the decision region of $\mathbf{c}_0$ is identical to that of the optimal ML decoder, as can be seen from Fig.~\ref{fig:DesionRegions_H1H2H3_LP_H3}.

\section{Finding the Minimal Weight Generators}
\label{sec:search for low weight pseudocodeword}

Inspired by the work of \cite{Chertkov-Polytope-of-Correct} we were encouraged to seek for a deterministic approach for finding the minimal weight generators. The heuristic method of \cite{Chertkov-Polytope-of-Correct} provides an excellent upper bound on the minimal pseudoweight, and it can be used for long and dense codes. It still lacks the certificate that the minimal weight generator that was found using this method is the minimal weight generator of a given parity check matrix. The number of iterations that are needed to reach a tight bound is also left open. The number of minimal weight generators is fundamental for obtaining a union bound, but the method of \cite{Chertkov-Polytope-of-Correct} can only estimate this number.

In this section we present a method for finding the minimal weight generator of a given parity check matrix, as well as the number of minimal weight generators.

\begin{prop}
  \label{prop:Pseudoweight invariant}
  (\!\! \cite{Kelley-Pseudocodewords-of-Tanner-graphs}) The pseudoweight is invariant under scaling.
  \begin{IEEEproof}
    We need to prove that $w_p^{AWGNC}(\alpha\mathbf{p})=w_p^{AWGNC}(\mathbf{p})$ where $\alpha$ is a real positive number, and $\mathbf{p} \in  \mathbb{R}^n$ is a pseudocodeword.
    \begin{equation}
      \label{eq_scaling}
      w_p^{AWGNC}(\alpha \mathbf{p})=
      \frac {(\sum_{i=1}^{n}\alpha p_i)^2}{\sum_{i=1}^{n}(\alpha p_i)^2}=
      \frac {\alpha^2(\sum_{i=1}^{n}p_i)^2}{\alpha^2\sum_{i=1}^{n}(p_i)^2}=w_p^{AWGNC}(\mathbf{p}).
    \end{equation}
  \end{IEEEproof}
\end{prop}

Given a binary parity check matrix $ H \in \mathbb{F}_2^{m \times n} $, the fundamental cone $ \mathcal{K} \buildrel \triangle \over = \mathcal{K}(H) $ is defined as the conic hull of the fundamental polytope $ P(H)$, and can be described by the following set of linear inequalities:
\begin{equation}
  \label{cone_eq}
  \mathcal{K}=\left\{\mathbf{p} \in \mathbb{R}^n \left |
      \begin{array}{l}
        p_i \geq 0 \; and\\
        \sum_{i=1, i\neq i'}^{n}h_{ji}p_i \geq h_{ji'}p_{i'}\\
        \forall \; 1 \leq i \leq n,\; \forall \; 1 \leq j \leq m
      \end{array}
    \right.
  \right\}
\end{equation}
where $ h_{ji} $ denotes the entry of H in the $j$th row and $i$th column.


\begin{cor}
  \label{cor:weights_on_an_edge}
  All points on an edge of the fundamental cone have the same pseudoweight.
  \begin{IEEEproof}
    Let $\mathbf{a}$ and $\mathbf{b}$ be two points on the edge $E_i \in \mathcal{K}$. From the definition of $\mathcal{K}$ it is clear that each edge is a ray with an endpoint at the origin. Being on the same ray, one can express $\mathbf{a} =  \alpha\mathbf{b} $ where $\alpha$ is a real positive number. From Proposition \ref{prop:Pseudoweight invariant} it follows that $\mathbf{a}$ and $\mathbf{b}$ have the same pseudoweight.
  \end{IEEEproof}
\end{cor}

Searching for the minimal weight generator is equivalent to searching for the minimal weight edge of the fundamental cone $\mathcal{K}$. We will now prove that for searching for the minimal weight edge one can bound $ \mathcal{K} $ and use the following polytope:

\begin{equation}
  \label{bounded_cone_eq}
  \mathcal{P}_\mathcal{K}=\left\{\mathbf{p} \in \mathbb{R}^n \left |
      \begin{array}{l}
        p_i \geq 0 \; and\\
        \sum_{i=1}^{n}p_i = a, \; a > 0\\
        \sum_{i=1, i\neq i}^{n}h_{ji}p_i \geq h_{ji'}p_{i'}\\
        \forall 1 \leq i \leq n, \forall 1 \leq j \leq m
      \end{array}
    \right.
  \right\}
\end{equation}
where the essence of the constant $ a \in \mathbb{R} $ is to increase the dynamic range of the problem and prevent scaling issues in optimization softwares.

\begin{prop}
  \label{prop:a_does_not_affect_the_wieght}
  The edges of $ \mathcal{K} $ and the vertices of $ \mathcal{P}_\mathcal{K} $ have the same weight distribution.
  \begin{IEEEproof}
    Let $E_i$ be an edge of $\mathcal{K}$ and let $\mathbf{p}$ be an arbitrary point on $E_i$. From (\ref{bounded_cone_eq}) it is clear that $E_i$ is also an edge of $ \mathcal{P}_\mathcal{K} $. Suppose $ \sum_{i=1}^{n}p_i~=~ b $, and let $ k = \frac{a}{b} $, then according to (\ref{eq_scaling}) the pseudocodeword $\mathbf{p'}=k\mathbf{p}$ has the same weight as $\mathbf{p}$ and the sum of its components is $a$;  Thus $ \mathbf{p'} $ is a vertex of $ \mathcal{P}_\mathcal{K} $.
    Now, let $E_j$ be an edge of $ \mathcal{P}_\mathcal{K} $. From the definition of $\mathcal{P}_\mathcal{K}$, $E_j$ has two endpoints: One in the origin and one in $\mathbf{\tilde{p}}$ for which $ \sum_{i=1}^{n}\tilde{p}_i = a $. Clearly, $ \mathcal{K} $ has an edge with an endpoint in the origin, that passes through $\mathbf{\tilde{p}}$ and goes off to infinity. According to Corollary \ref{cor:weights_on_an_edge} the pseudoweight of this edge in $ \mathcal{K} $ is the same as the vertex $\mathbf{\tilde{p}} \in \mathcal{P}_\mathcal{K}$, which completes the proof.
  \end{IEEEproof}
\end{prop}

The problem of finding the minimal weight in an AWGN channel becomes:



\begin{equation}
  \label{minimal_weight_generator_eq}
  w_{p_{min}} = \min_{\mathbf{p} \in \mathcal{P}_\mathcal{K}} w_p(\mathbf{p}) =
  \min_{\mathbf{p} \in \mathcal{P}_\mathcal{K}} \frac {(\sum_{i=1}^{n}p_i)^2}{\sum_{i=1}^{n}(p_i)^2} =
  \min_{\mathbf{p} \in \mathcal{P}_\mathcal{K}} \frac {a^2}{\sum_{i=1}^{n}(p_i)^2}
\end{equation}
where the last equation follows from the definition of $ \mathcal{P}_\mathcal{K} $. Being a constant, $a$ does not affect the minimization process, thus instead of solving (\ref{minimal_weight_generator_eq}) one may consider solving the following simpler maximization problem:
\begin{equation}
  \label{maximization_minimal_weight_eq}
  \max_{\mathbf{p} \in \mathcal{P}_\mathcal{K}} \sum_{i=1}^{n}(p_i)^2.
\end{equation}
The minimal pseudoweight $w_{p_{min}}$ is simply the division of $a^2$ by the optimal solution of (\ref{maximization_minimal_weight_eq}).

The maximization problem of (\ref{maximization_minimal_weight_eq}) is non-convex and may have several local maxima. Algorithms for solving such problems are termed \textit{Global Optimization} and are able to find the global solution in the presence of multiple local solutions.





Global optimization algorithms are usually divided into deterministic and probabilistic approaches. The solution of a deterministic approach is guaranteed to be the global solution, or at least a local solution which differs from the global solution by less than a given $\epsilon > 0$.
Probabilistic algorithms require a shorter runtime compares to deterministic ones, but their solution may not be the global optimum.

\begin{table} [b!]
  \centering
  \begin{tabular}{|l|l|l|}
    \hline
    \textbf{Code} & $ \mathbf{ w_{p_{min}} } $ & $ \mathbf { N_{p_{min}} } $  \\
    \hline
    \hline
    Hamming [15, 11, 3]  & 3.0 & 127\\
    \hline
    Hamming [31, 26, 3]  & 3.0 & 1185\\
    \hline
    BCH [31, 21, 5]  & 3.0 & 6\\
    \hline
    BCH [63, 45, 7]  & 3.0 & 1\\
    \hline
    BCH [63, 39, 9]  & 3.299176 & 1\\
    \hline
    BCH [63, 36, 11]  & 3.2 & 83\\
    \hline
    BCH [127, 113, 5]  & 3.0 & 134\\
    \hline
    BCH [127, 64, 21]  & 3.0 & 2\\
    \hline
    BCH [255, 131, 37]  & 3.33333 & 9\\
    \hline
    Tanner [155, 64, 20]  & 16.403683 & 465\\
    \hline
  \end{tabular}
  \caption{Minimal-weight and number of minimal-weight generators}
  \label{table:minimal_weight_generators}
\end{table}

An efficient deterministic approach for solving global optimization problems is the \textit{Branch and Bound} algorithm. The algorithm relies on the existence of a convex relaxation of the original problem \cite{Twarmalani-Sahinidis-2002b},  whose optimal solution provides a lower bound on the solution of the original problem.

The simplest probabilistic global optimization algorithm is the \textit{Multistart} algorithm, which uses a local algorithm starting from several points distributed over the whole optimization region. The local optimum with the best objective value is taken as the global solution.

Some global optimization softwares such as \textit{BARON} (Branch And Reduce Optimization Navigator) \cite{Sahinidis-Baron} can also provide the $k$-best solutions, or all local solutions, which guarantee not only finding the minimal weight, but also the distribution of the minimal weight generators.

Being deterministic and efficient, BARON was chosen as the global optimization software. The minimal weight along with the number of minimal weight generators $ \mathbf { N_{p_{min}} }$ for several selected codes are presented in Table \ref{table:minimal_weight_generators}. We performed short cycles reduction \cite{Halford-Random-Iterative} on the BCH codes to improve their performance under iterative decoding. The minimal pseudoweight for the [155,~64,~20] Tanner code \cite{Tanner-LDPC-Block} presented in \cite{Chertkov-Efficient-Pseudo-Codeword-Search-Algorithm} is $d_{LP} \approx 16.4037$, which is similar to our results. Notice that the LDPC Tanner code has a minimal pseudoweight much higher than all the tested HDPC codes. Although short cycles were removed from the BCH codes, we were not able to increase the minimal pseudoweight beyond $3\frac{1}{3}$. It is interesting to develop new methods for increasing the minimal pseudoweight of a given dense parity check matrix without adding redundant parity checks.

\section{LP Union Bound}
\label{sec:LP Union Bound}

A union bound for LP decoding was mentioned in \cite{Chertkov-Reducing}, \cite{Smarandache-Pseudo-codeword-analysis} and \cite{Skachek-Lower-bounds-on-the-minimum-pseudodistance}, but a full characterization of such a bound was not provided. In this section we examine the [8, 4, 4] Extended Hamming code and the [24, 12, 8] extended Golay code, and present why calculation of such a bound is not an easy task.

The ML union bound \cite{Barry-Digital-Communication} for the case where the all-zero codeword $ \mathbf{s}_0 $ is transmitted is

\begin{equation}
  \label{ML_ub}
  Pr[error|\mathbf{s}_0] \le \sum_{i=1}^{M-1}Q\left(\frac{d_{0i}}{2\sigma}\right)
\end{equation}\\
where $ M $ is the number of signals and $ d_{0i} $ is the distance between $ \mathbf{s}_0 $ and $ \mathbf{s}_i $. Clearly, (\ref{ML_ub}) can be very loose in case the individual events are not disjoint. For large SNRs the union bound of (\ref{ML_ub}) can be approximated by including only the dominating terms:

\begin{equation}
  \label{ML_approx_ub}
  Pr[error|\mathbf{s}_0] \approx N_{min}Q\left(\frac{d_{min}}{2\sigma}\right)
\end{equation}\\
where $ {N_{min}} $ is the number of nearest neighbors of the transmitted signal $ \mathbf{s}_0 $. We can no longer assume that (\ref{ML_approx_ub}) is an upper bound, since we've neglected positive terms from (\ref{ML_ub}).

Assuming that the all-zero codeword is transmitted, the error probability over the fundamental polytope is equal to that over the fundamental cone \cite{Siegel-Relaxation-bounds}; Thus, a union bound for the LP decoder can be formulated from (\ref{ML_ub}) as follows:
\begin{equation}
  \label{LP_ub}
  Pr[error|\mathbf{s}_0] \le \sum_{i=1}^{N_p-1}Q\left(\frac{w_{p_i}}{2\sigma}\right)
\end{equation}
where $ N_p $ is the number of generators and $ w_{p_i} $ is the pseudoweight of generator $ i $. Similarly, we can obtain an approximation for large SNRs:
\begin{equation}
  \label{LP_approx_ub}
  Pr[error|\mathbf{s}_0 \; transmitted] \approx N_{p_{min}}Q\left(\frac{w_{p_{min}}}{2\sigma}\right)
\end{equation}
where $ N_{p_{min}} $ is the number of minimal weight generators, and $ w_{p_{min}} $ is the minimal pseudoweight.

In the following we will use the parity check matrices of (\ref{H1}), (\ref{H2}) and (\ref{H3}) for the Extended [8, 4, 4] Hamming code and (\ref{HG}), (\ref{HG'}) for the [24, 12, 8] extended Golay code.

\begin{equation}
  \label{HG}
  H_G=\left(
    \begin{array}{p{0.15cm}p{0.15cm}p{0.15cm}p{0.15cm}p{0.15cm}p{0.15cm}p{0.15cm}p{0.15cm}p{0.15cm}p{0.15cm}p{0.15cm}p{0.15cm}p{0.15cm}p{0.15cm}p{0.15cm}p{0.15cm}p{0.15cm}p{0.15cm}p{0.15cm}p{0.15cm}p{0.15cm}p{0.15cm}p{0.15cm}p{0.15cm}}
      1&0&0&1&1&0&1&0&1&1&1&1&0&0&0&0&0&1&0&1&0&0&1&1\\
      1&1&0&0&1&1&0&1&0&1&1&1&1&0&0&0&0&0&1&0&1&0&0&1\\
      0&1&1&0&0&1&1&0&1&0&1&1&1&1&0&0&0&0&0&1&0&1&0&1\\
      0&0&1&1&0&0&1&1&0&1&0&1&1&1&1&0&0&0&0&0&1&0&1&1\\
      1&0&0&1&1&0&0&1&1&0&1&0&1&1&1&1&0&0&0&0&0&1&0&1\\
      0&1&0&0&1&1&0&0&1&1&0&1&0&1&1&1&1&0&0&0&0&0&1&1\\
      1&0&1&0&0&1&1&0&0&1&1&0&1&0&1&1&1&1&0&0&0&0&0&1\\
      0&1&0&1&0&0&1&1&0&0&1&1&0&1&0&1&1&1&1&0&0&0&0&1\\
      0&0&1&0&1&0&0&1&1&0&0&1&1&0&1&0&1&1&1&1&0&0&0&1\\
      0&0&0&1&0&1&0&0&1&1&0&0&1&1&0&1&0&1&1&1&1&0&0&1\\
      0&0&0&0&1&0&1&0&0&1&1&0&0&1&1&0&1&0&1&1&1&1&0&1\\
      1&1&1&1&1&1&1&1&1&1&1&1&1&1&1&1&1&1&1&1&1&1&1&1\\
    \end{array}
  \right)
\end{equation}

\begin{equation}
  \label{HG'}
  H_{G'}=\left(
    \begin{array}{p{0.15cm}p{0.15cm}p{0.15cm}p{0.15cm}p{0.15cm}p{0.15cm}p{0.15cm}p{0.15cm}p{0.15cm}p{0.15cm}p{0.15cm}p{0.15cm}p{0.15cm}p{0.15cm}p{0.15cm}p{0.15cm}p{0.15cm}p{0.15cm}p{0.15cm}p{0.15cm}p{0.15cm}p{0.15cm}p{0.15cm}p{0.15cm}}
      1&0&0&1&1&0&1&0&1&1&1&1&0&0&0&0&0&1&0&1&0&0&1&1\\
      0&1&0&0&1&0&0&0&1&1&0&0&0&1&0&0&0&0&1&0&1&0&0&1\\
      1&1&1&0&0&0&1&1&0&0&0&0&0&0&0&0&0&0&0&1&0&1&0&1\\
      0&0&0&1&0&0&1&0&0&0&0&1&1&0&1&0&1&0&0&0&0&1&1&0\\
      1&0&0&0&0&1&0&1&1&0&1&1&1&1&0&0&0&0&0&0&0&0&0&0\\
      0&0&0&0&0&1&0&0&0&0&0&1&0&0&1&1&1&0&1&0&1&0&1&0\\
      1&0&1&0&0&1&1&0&0&1&1&0&1&0&1&1&1&1&0&0&0&0&0&1\\
      0&0&1&1&0&1&0&1&1&0&0&0&1&0&0&1&1&1&1&1&0&1&0&0\\
      0&0&0&1&1&1&0&0&0&0&0&1&0&0&1&1&0&0&0&0&0&1&0&1\\
      0&0&1&0&0&0&0&1&0&1&0&0&0&1&0&0&1&0&0&0&1&1&0&1\\
      0&0&1&0&1&0&1&1&0&0&1&0&0&0&1&0&0&0&1&1&0&0&0&0\\
      1&1&0&1&0&1&1&0&0&1&1&0&0&1&0&1&0&0&0&0&1&1&1&0\\
    \end{array}
  \right)
\end{equation}

The parity check matrices for the [24,~12,~8] extended Golay code were introduced by Halford in \cite{Halford-thesis}. The parity check matrix in (\ref{HG'}) was obtained by applying short cycles reduction \cite{Halford-Random-Iterative} on (\ref{HG}). The Tanner graph representation of $H_G$ contains 1,551 4-cycles and 65,632 6-cycles, while there are 295 4-cycles and 6,204 6-cycles in the representation of $H_{G'}$. It was shown in \cite{Halford-thesis} that the message passing algorithm using $H_{G'}$ outperforms the one using $H_G$ by more than 1~dB.

From Section \ref{sec:the affect of minimal weight pseudocodewords on lp decoding}, it is clear why having a tight LP upper bound is a complicated task: In the low SNR regime the bound may be very loose if not scaling each minimal weight pseudocodeword by its contribution to the error probability. In the high SNR regime, especially for medium and long codes, the minimal weight pseudocodewords may have low-volume decision regions, thus may have a negligible effect on the performance of the LP decoder. A similar phenomenon was observed in \cite{Amrani-ounded-distance-decoding}, \cite{Fishler-Geometrical} for \textit{pseudo nearest neighbors} in bounded-distance decoding algorithms. In \cite{Fishler-Geometrical} Fishler et al. derived an approximated probability ratio between the error contribution of a non-codeword neighbor and a codeword nearest neighbor. The ratio was calculated based on the ratio between the volumes of the decision regions of the two competitive neighbors. Incorporating this ratio yielded a better approximation for an upper bound.

\begin{figure}
  \centering

  \subfigure[{[8, 4, 4] extended Hamming code}]{
    \includegraphics[scale=0.6]{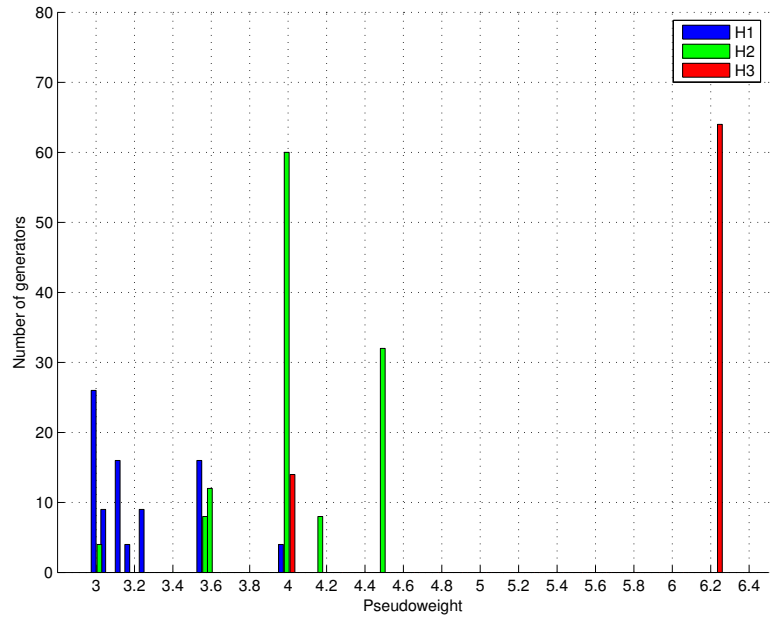}
    \label{fig:DesionRegions_H1H2H3_generators_histogram}}

  \subfigure[{[24, 12, 8] extended Golay code}]{
    \includegraphics[scale=0.62]{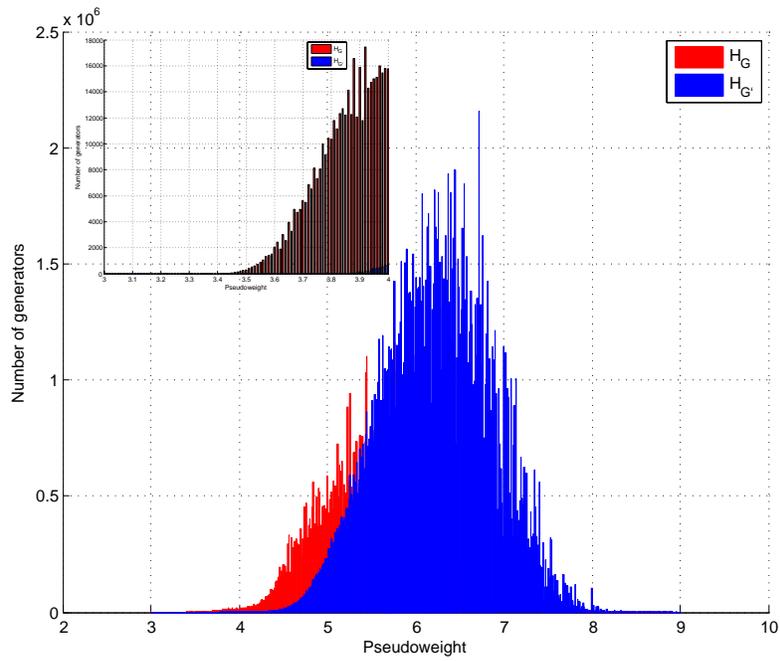}
    \label{fig:DesionRegions_golay_generators_histogram}}

  \caption{Weight distribution of the $[8, 4, 4]$ extended Hamming code and the $[24, 12, 8]$ extended Golay code.}
  \label{fig:DesionRegions_generators_histogram}
\end{figure}

Fig.~\ref{fig:DesionRegions_generators_histogram} presents the generators' weight distribution of the [8, 4, 4] extended Hamming code and the [24,~12,~8] extended Golay code. Clearly, using (\ref{LP_ub}) as an upper bound for the LP decoder will result an error probability much higher than unity, which makes the bound useless. The reason is that the large number of generators leads to many joint events, which make the bound very loose. Strictly speaking, while the [24,~12,~8] extended Golay code has 4096 codewords, there are 91,113,330 and 231,146,334 generators for $H_G$ and $H_{G'}$, respectively. It was mentioned above that $H_{G'}$ was obtained from $H_G$ by applying short cycles reduction. Notice that not only $H_{G'}$ has a higher minimal pseudoweight, but its entire pseudoweight spectra is centered to the right of the one of $H_G$, as illustrated in Fig.~\ref{fig:DesionRegions_golay_generators_histogram}.

\begin{table} [h]
  \centering
  \begin{tabular}{|c|c|c|c|}
    \hline
    \textbf{Code} & \textbf{Parity Check Matrix} & $ \mathbf{ w_{p_{min}} } $ & $ \mathbf { N_{p_{min}} }$  \\
    \hline
    \hline
    Extended Hamming [8, 4, 4] & $H_1$ & 3.0 & 26\\
    \cline{2-4}
    & $H_2$ & 3.0 & 4\\
    \cline{2-4}
    & $H_3$ & 4.0 & 14\\
    \hline
    Extended Golay [24, 12, 8] & $H_G$ & 3.0 & 2\\
    \cline{2-4}
    & $H_{G'}$ & 3.6 & 2\\
    \hline
  \end{tabular}
  \caption{Minimal-weight generators  of the $[8, 4, 4]$ extended Hamming code and the $[24, 12, 8]$ extended Golay code}\label{table:Minimal_weight_generators}
\end{table}

\begin{figure}
  \centering

  \subfigure[{[8, 4, 4] extended Hamming code}]{
    \includegraphics[scale=0.6]{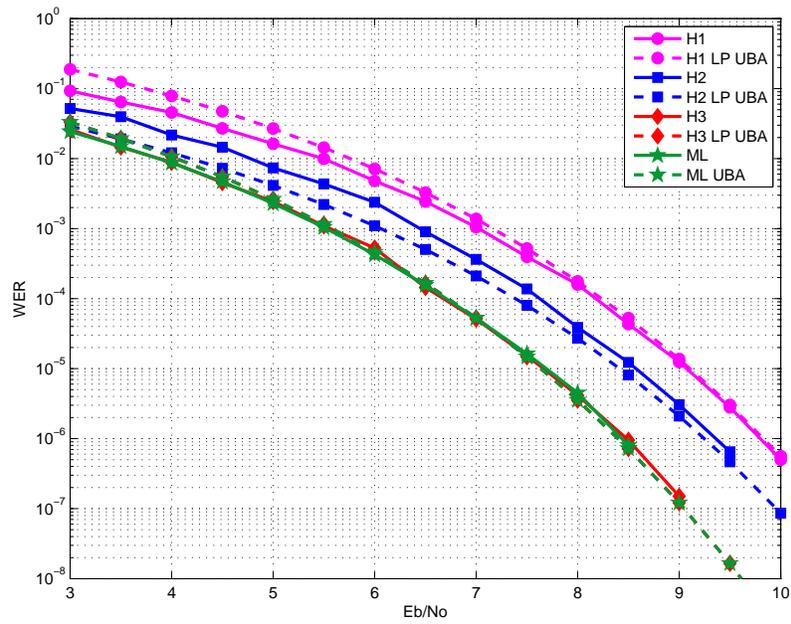}
    \label{fig:DesionRegions_LP_UB_h1h2h3}
  }

  \subfigure[{[24, 12, 8] extended Golay code}]{
    \includegraphics[scale=0.6]{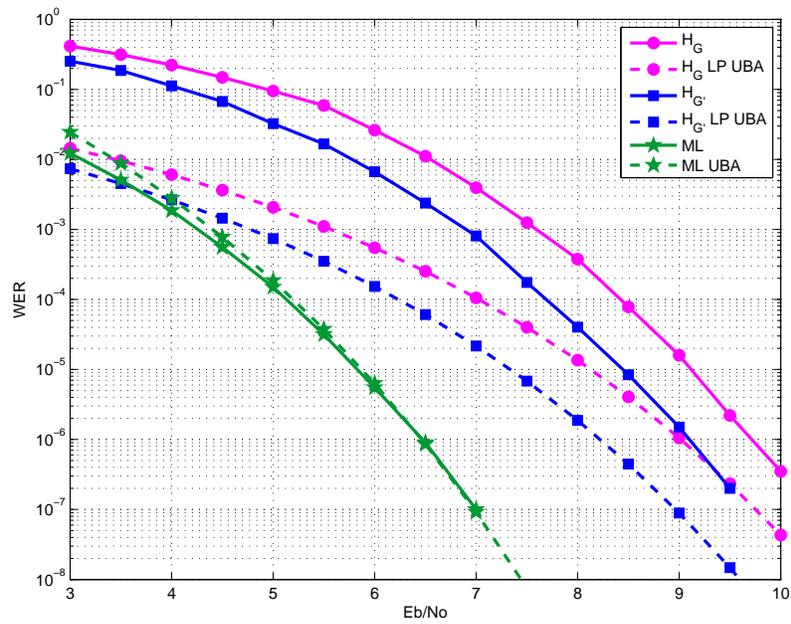}
    \label{fig:DesionRegions_LP_UB_golay}
  }

  \caption{LP UBA for different representations of the $[8, 4, 4]$ extended Hamming code and the $[24, 12, 8]$ extended Golay code.}
  \label{fig:DesionRegions_LP_UB}
\end{figure}

Table \ref{table:Minimal_weight_generators} presents the number of minimal-weight generators of the aforementioned parity check matrices. The pseudoweights from Table \ref{table:Minimal_weight_generators} were used to calculate the LP Union Bound Approximation (UBA) of (\ref{LP_approx_ub}). The results are presented in Fig.~\ref{fig:DesionRegions_LP_UB}. For clarity the actual error probabilities are also presented. The performance of the LP decoder for the chosen matrices are correlated with the generators' weight distribution that was presented in Fig.~\ref{fig:DesionRegions_generators_histogram}. From Fig \ref{fig:DesionRegions_LP_UB_h1h2h3} one can see that the suggested UBA is tight for $ H_1 $ and $ H_3 $, but inaccurate for $ H_2 $. The UBA is much worse for the case of the extended Golay code, as presents in Fig.~\ref{fig:DesionRegions_LP_UB_golay}. In this case the UBA does not reflect the actual behavior of the decoder, but rather presents a too-optimistic behavior. The reason is due to the fact that there are only two minimal-weight generators for both representations, whereas the ML UBA (\ref{ML_approx_ub}) employs 759 nearest neighbors. Notice that there are 91,113,326 and 230,918,045 generators of pseudoweight lower than $d_{min}$ for $H_G$ and $H_{G'}$, respectively. Had we considered all the generators having pseudoweight lower than $d_{min}$ in (\ref{LP_approx_ub}), we would have gained an LP UBA higher than unity. The LP UBA does not have a coherent behavior, i.e. the bound can be sometimes below the actual performance curve, which again disqualifies it as an upper bound or as an approximation.

The pseudoweight and the weight-distribution are not enough for implementing a tight LP upper bound. A tight and accurate bound must take the volume of the decision regions into account. As was presented, some low-weight pseudocodewords which have small volume have also small effect on the decoder's performance, but are very dominant in the equation of the LP union-bound. A tight LP union-bound must have a distinction between codewords and different types of pseudocodewords.

\section{Conclusion}
\label{sec:Conclusion}

The decision regions of the LP decoder were studied and compared to those of the BP and ML decoders. We showed that both BP and LP are not bounded distance decoders. The different effect of minimal-weight pseudocodewords on the performance of the LP decoder was examined. Global optimization was presented as a method for finding the minimal pseudoweight, as well as the number of minimal weight generators. An LP union bound was presented, along with an explanation of why having the pseudoweight spectra is not sufficient for finding a tight bound. Finding the ratio between the error contribution of a nearest pseudocodeword and a nearest codeword will tighten the union bound, and is left open for further research.


\begin{thebibliography}{10}
  \providecommand{\url}[1]{#1}
  \csname url@samestyle\endcsname
  \providecommand{\newblock}{\relax}
  \providecommand{\bibinfo}[2]{#2}
  \providecommand{\BIBentrySTDinterwordspacing}{\spaceskip=0pt\relax}
  \providecommand{\BIBentryALTinterwordstretchfactor}{4}
  \providecommand{\BIBentryALTinterwordspacing}{\spaceskip=\fontdimen2\font plus
    \BIBentryALTinterwordstretchfactor\fontdimen3\font minus
    \fontdimen4\font\relax}
  \providecommand{\BIBforeignlanguage}[2]{{%
      \expandafter\ifx\csname l@#1\endcsname\relax
      \typeout{** WARNING: IEEEtran.bst: No hyphenation pattern has been}%
      \typeout{** loaded for the language `#1'. Using the pattern for}%
      \typeout{** the default language instead.}%
      \else
      \language=\csname l@#1\endcsname
      \fi
      #2}}
  \providecommand{\BIBdecl}{\relax}
  \BIBdecl

\bibitem{Kschischang-Factor-graphs}
  F.~R. Kschischang, B.~J. Frey, and H.-A. Loeliger, ``Factor graphs and the
  sum-product algorithm,'' \emph{{IEEE} Trans. Inf. Theory}, vol.~47, no.~2,
  pp. 498--519, Feb. 2001.

\bibitem{Vontobel-On-the-relationship}
  P.~O. Vontobel and R.~Koetter, ``On the relationship between linear programming
  decoding and min-sum decoding,'' in \emph{Proc. IEEE International Symposium
    on Information Theory and its Applications}, Parma, Italy, Oct. 2004.

\bibitem{Vontobel-Graph-cover-decoding}
  \BIBentryALTinterwordspacing
  P.~O. Vontobel and R.~Koetter, ``Graph-cover decoding and finite-length analysis of message-passing
  iterative decoding of ldpc codes,'' 2005. [Online]. Available:
  \url{http://www.arxiv.org/abs/cs.IT/0512078}
  \BIBentrySTDinterwordspacing

\bibitem{Feldman-Using-linear-programming}
  J.~Feldman, M.~J. Wainwright, and D.~R. Karger, ``Using linear programming to
  decode binary linear codes,'' \emph{{IEEE} Trans. Inf. Theory}, vol.~51,
  no.~1, pp. 954--972, Jan. 2005.

\bibitem{Siegel-Relaxation-bounds}
  \BIBentryALTinterwordspacing
  P.~Chaichanavong and P.~H. Siegel, ``Relaxation bounds on the minimum
  pseudoweight of linear block codes,'' in \emph{Proc. IEEE International
    Symposium on Information Theory}, no. 805-809, Adelaide, Australia, Sep. 4-9
  2005. [Online]. Available: \url{http://www.arxiv.org/abs/cs.IT/0508046}
  \BIBentrySTDinterwordspacing

\bibitem{Kashyap-decomposition-theory}
  N.~Kashyap, ``A decomposition theory for binary linear codes,'' \emph{{IEEE}
    Trans. Inf. Theory}, vol.~54, no.~7, pp. 3035--30\,589, July. 2008.

\bibitem{Vontobel-Lower-bounds}
  P.~O. Vontobel and R.~Koetter, ``Lower bounds on the minimum pseudo-weight of
  linear codes,'' in \emph{IEEE International Symposium on Information Theory},
  Chicago, IL, USA., June 27-July 2 2004, p.~70.

\bibitem{Wiberg-Codes-and-Decoding}
  N.~Wiberg, ``Codes and decoding on general graphs,'' Ph.D. dissertation,
  Linkoping University, Linkoping, Sweden, 1996.

\bibitem{Kelley-Pseudocodewords-of-Tanner-graphs}
  C.~Kelley and D.~Sridhara, ``Pseudocodewords of tanner graphs,'' \emph{{IEEE}
    Trans. Inf. Theory}, vol.~53, no.~11, pp. 4013--4038, Nov. 2007.

\bibitem{Frey-Signal-space-characterization}
  B.~J. Frey, R.~Koetter, and A.~Vardy, ``Signal space characterization of
  iterative decoding,'' \emph{{IEEE} Trans. Inf. Theory}, vol.~47, no.~2, pp.
  766--781, Feb. 2001.

\bibitem{Forney-On-the-effective-weights}
  G.~D. Forney, Jr., R.~Koetter, F.~R. Kschischang, and A.~Reznik, ``On the
  effective weights of pseudocodewords for codes defined on graphs with
  cycles,'' in \emph{Codes, Systems and Graphical Models}, ser. {IMA} Volumes
  in Mathematics and Its Applications.\hskip 1em plus 0.5em minus 0.4em\relax
  New York/Minneapolis: Springer-Verlag, Nov. 1998, vol. 123, pp. 101--112.

\bibitem{Chertkov-Efficient-Pseudo-Codeword-Search-Algorithm}
  M.~Chertkov and M.~Stepnov, ``An efficient pseudo-codeword-search algorithm for
  linear programming decoding of ldpc codes,'' \emph{{IEEE} Trans. Inf.
    Theory}, vol.~54, no.~4, pp. 1514--1520, Apr. 2008.

\bibitem{Vontobel-On-the-minimal-pseudo-codewords}
  \BIBentryALTinterwordspacing
  P.~O. Vontobel, R.~Smarandache, N.~Kiyavash, J.~Teutsch, and D.~Vukobratovic,
  ``On the minimal pseudo-codewords of codes from finite geometries,'' in
  \emph{Proc. IEEE International Symposium on Information Theory}, no. 980-984,
  Adelaide, Australia, Sep. 4-9 2005. [Online]. Available:
  \url{http://www.arxiv.org/abs/cs.IT/0508019}
  \BIBentrySTDinterwordspacing

\bibitem{Chertkov-Reducing}
  \BIBentryALTinterwordspacing
  M.~Chertkov, ``Reducing the error floor,'' in \emph{Proc. Information Theory
    Workshop}, no. 230 - 235, Lake Tahoe, CA, USA, Sep.2-6 2007. [Online].
  Available: \url{http://arxiv.org/abs/0706.2926v1}
  \BIBentrySTDinterwordspacing

\bibitem{Halford-Random-Iterative}
  T.~R. Halford and K.~M. Chugg, ``Random redundant iterative soft-in soft-out
  decoding,'' \emph{{IEEE} Trans. Commun.}, vol.~56, no.~4, pp. 513--517, Apr.
  2008.


\bibitem{Chertkov-Polytope-of-Correct}
  \BIBentryALTinterwordspacing
  M.~Chertkov and M.~Stepanov, ``Polytope of correct (linear programming)
  decoding and low-weight pseudo-codewords,'' Feb. 2011. [Online]. Available:
  \url{http://arxiv.org/abs/1102.3902}
  \BIBentrySTDinterwordspacing

\bibitem{Amrani-ounded-distance-decoding}
  O.~Amrani and Y.~Be'ery, ``Bounded-distance decoding: algorithms, decision
  regions, and pseudo nearest-neighbors,'' \emph{{IEEE} Trans. Inf. Theory},
  vol.~44, no.~7, pp. 3072--3082, Nov. 1998.

\bibitem{Fishler-Geometrical}
  E.~Fishler, O.~Amrani, and Y.~Be'ery, ``Geometrical and performance analysis of
  gmd and chase decoding algorithms,'' \emph{{IEEE} Trans. Inf. Theory},
  vol.~45, no.~5, pp. 1406--1422, 1999.

\bibitem{Twarmalani-Sahinidis-2002b}
  N.~V. Sahinidis and M.~Twarmalani, \emph{Convexification and Global
    Optimization in Continuous and Mixed-Integer Nonlinear Programming}.\hskip
  1em plus 0.5em minus 0.4em\relax Kluwer Academic Publishers Group, 2002.

\bibitem{Sahinidis-Baron}
  N.~V. Sahinidis, ``{BARON}: A general purpose global optimization software
  package,'' \emph{Journal of Global Optimization}, vol.~8, pp. 201--205, 1996.

\bibitem{Tanner-LDPC-Block}
R.~M. Tanner, D.~Sridhara, A.~Sridharan, T.~E. Fuja, and D.~J. Costello, ``Ldpc
  block and convolutional codes based on circulant matrices,'' \emph{{IEEE}
  Trans. Inf. Theory}, vol.~50, pp. 2966--2984, 2004.

\bibitem{Smarandache-Pseudo-codeword-analysis}
  R.~Smarandache and P.~O. Vontobel, ``Pseudo-codeword analysis of tanner graphs
  from projective and euclidean planes,'' \emph{{IEEE} Trans. Inf. Theory},
  vol.~53, no.~7, pp. 2376--2393, Jul. 2007.

\bibitem{Skachek-Lower-bounds-on-the-minimum-pseudodistance}
  V.~Skachek and M.~F. Flanagan, ``Lower bounds on the minimum pseudodistance for
  linear codes with q-ary PSK modulation over AWGN,'' in \emph{Proc. 5th
    International Symposium on Turbo Codes and Related Topics}, Lausanne,
  Switzerland, Sep. 1-5 2008.

\bibitem{Barry-Digital-Communication}
  J.~R. Barry, E.~A. Lee, and D.~G. Messerschmitt, \emph{Digital Communication},
  3rd~ed.\hskip 1em plus 0.5em minus 0.4em\relax Springer, 2004.

\bibitem{Halford-thesis}
T.~R. Halford, ``The extraction and complexity limits of graphical models for
  linear codes,'' Ph.D. dissertation, Los Angeles, CA, USA, 2007.

\end{thebibliography}

\end{document}